# Complex coalitions:
## political alliances across relational contexts

Malkamäki, Arttu[1,2]; Chen, Ted Hsuan Yun[3]; Gronow, Antti[1]; Kivelä, Mikko[4]; Vesa, Juho[1]; Ylä-Anttila, Tuomas[1]

[1]Faculty of Social Sciences, University of Helsinki, Finland; [2]Department of Communication (visiting scholar), Stanford University, United States; [3]Department of Environmental Science and Policy, George Mason University, United States; [4]Department of Computer Science, Aalto University, Finland

**Abstract**

Coalitions are central to politics, including government formation, international relations, and public policy. Coalitions emerge when actors engage one another across multiple relational contexts, but existing literature often approaches coalitions in singular contexts. We introduce *complex coalitions*, a theoretical-methodological framework that emphasises the relevance of multiple contexts and cross-context dependencies in coalition politics. We also implement tools to statistically infer such coalition structures using multilayer networks. To demonstrate the usefulness of our approach, we compare coalitions among Finnish organisations engaging in climate politics across three contexts: resource coordination, legacy media discourse, and social media communication. We show that considering coalitions as complex and accounting for cross-context dependencies improves the empirical validity of coalition studies. In our case study, the three contexts represent complementary, but not congruent, channels for enacting coalitions. In conclusion, we argue that the complex coalitions approach is useful for advancing understanding of coalitions in different political realms.

## 1 Introduction

Coalitions are social mechanisms for actors engaging in politics to organise joint action for mutual gain. The goals, composition, longevity, success, and scholarly definitions of coalitions vary, ranging from coalitions of political parties that form a government, military alliances, and social movement coalitions around protest events to advocacy coalitions of actors who join forces to steer public policy (Ecker and Meyer 2020; Leeds 2003; McClendon 2014; Weible, Ingold, Nohrstedt, Henry, et al. 2020). As coalitions drive collective action across various political phenomena, understanding how coalitions emerge and contribute to diverse outcomes, including political divides (Fortunato and Stevenson 2013), policy deadlocks (Farrell 2016), party position changes (Karol 2009), governance cultures (Johan and Lindahl 2021), international sanctions (Hagen and Schneider 2021), or even reshuffling of the global order (Roberts 2019), is core to the study of politics.

The basic task for researchers aiming to understand such phenomena is inferring coalition boundaries, i.e. identifying which actors belong to which coalitions. Common to the literature is that coalitions are predicated upon ties that occur among actors. For example, social movement coalitions often emerge from ties based on conducive organisational structures and shared identities (Van Dyke and Amos 2017), military alliances focus on collective provision of security (e.g., joint reconnaissance) based on shared threats (Leeds 2003; Weitsman 1997), and actors who wish to change or defend particular policies typically coordinate activities based on shared beliefs (Sabatier 1998). Thus, coalitions emerge and are inferred by scholars from interactions among political actors, which stem from various actor-level, group-level, and system-level factors and dynamics (Metz and Brandenberger n.d.).

The political network approach and its corresponding network analysis methods, which deal with ties among political units, have become effective ways of studying social organisation in politics (Hafner-Burton, Kahler, and Montgomery 2009; Siegel 2009). To infer coalition boundaries, scholars are increasingly using algorithms to understand how groups emerge based on how ties between actors either cluster together or repulse from one another. Such network partitioning algorithms have been applied to infer, for example, cross-hemisphere coalitions of non-governmental organisations (Cheng, Wang, Ma, and Murdie 2021), cross-identity coalitions of social movements (Park 2008), rival groups of organisations engaging in land use conflict (Malkamäki, Ylä-Anttila, Brockhaus, Toppinen, et al. 2021), as well as bipartisan "shadow" coalitions of legislators and interest groups convening around parties in the US Congress (Aref and Neal 2021; Desmarais, La Raja, and Kowal 2015). Besides exposing coalition boundaries, scholars often analyse coalition dynamics to explain other outcomes, such as policy change or stability of the political system.

Yet, coalitions are usually treated as emerging from within singular relational contexts, and accordingly inferred from networks containing only a single type of a tie (e.g., communication) or those collapsing different types of ties into a single type (e.g., the intersection of communication and expression of trust becomes collaboration). As recent work on the generative processes underlying political networks indicates (Chen 2021), the ties giving rise to specific types of coalitions are, however, likely to unfold not only within, but across multiple relational contexts. For example, actors may interact simultaneously on social media and by holding joint press events (Chamberlain, Spezzano, Kettler, and Dit 2021; Desmarais,

Moscardelli, Schaffner, and Kowal 2015). The problem with focusing on ties in singular contexts is twofold. First, different contexts constrain tie formation differently and capture different strategic considerations by the actors (e.g., due to varying transaction costs). Second, ties in different contexts hardly occur independently of one another (e.g., information campaigning on social media hardly occurs in isolation from relying on the same advisor). Failure to account for variation and dependencies across contexts risks producing incomplete or misleading findings about coalitions, and much of politics.

We introduce *complex coalitions*, a theoretical-methodological framework that explicitly recognises coalitions as emergent, complex phenomena. We argue that understanding coalitions requires taking stock of the heterogeneity of ties across both modal (i.e., different types of ties) and temporal (i.e., ties at different times) contexts instead of focusing on singular types of ties or pooling all ties together. Coalitions that emerge from different contexts should also be taken as interdependent, exhibiting some degree of mutual reinforcement, instead of simply existing alongside one another. In addition to our theoretical contributions, we implement novel methodological tools for the study of coalitions, drawing from the notion of community structure in "multilayer" networks (Kivelä et al. 2014), that allow scholars to infer valid complex coalitions by incorporating ties across multiple contexts and accounting for cross-context dependencies. Unlike earlier work in political science, notably the study by Cranmer, Menninga, and Mucha (2015) concerning the fractionalisation of economic and political ties in world politics, we implement generative models of multilayer networks that include community structure in the model description and enable coalition inference according to the probability of the data under the model parameters (Pamfil, Howison, Lambiotte, and Porter 2019).

To illustrate how our multilayer approach performs against the conventional "monolayer" approach, we use eight years of data on climate politics in Finland, 2013-2020, focusing on ties occurring in three contexts: resource coordination, discursive alignment of political statements in legacy media, and endorsement of political content on social media. To demonstrate the need to consider coalitions as complex, we ask: *Do actors behave, and coalitions manifest, differently across relational contexts, and if so, why?*

In the second section, we present the complex coalitions approach and argue the need to consider multiple types of ties when inferring coalitions. Because actors tend to form ties strategically across contexts, focusing on any singular type of interaction risks missing important interdependencies. In the third section, we show how our theoretical arguments and methodological contributions work in the case of Finnish climate politics. We show that a) accounting for cross-context dependencies improves model fit, b) different relational contexts from which coalitions emerge exhibit dynamics that are largely complementary but not entirely congruent, and c) the complex coalitions approach provides a holistic view of phenomena such as power and polarisation in politics. Finally, we discuss our contributions considering the existing literature, as well as the generalisability, advantages, and limitations of the complex coalitions approach for the study of politics.

## 2 Complex coalitions

As attaining and maintaining political power is rarely possible without powerful allies, forming ties to other actors usually pays off (Boix and Svolik 2013; Mahoney and Baumgartner 2015). Coalitions, then, emerge from the intention to counter rival concentrations of power by forming ties with other actors. Identifying coalition boundaries and analysing coalition dynamics entails the task of observing all relevant ties among all relevant actors and inferring coalitions by recognising patterns among them.

Like most political phenomena, coalitions are emergent. The notion of emergence is associated with the behaviour of complex systems. It refers to relatively simple and local interactions (e.g., communication) among various units of the system causing complex and systemic behaviour (e.g., cohesion) that is often qualitatively different from the inputs (Morrison 2015). As collections of edges (i.e., interactions) and vertices (i.e., units), networks are ideal for studying emergent phenomena. Empirical networks typically host large-scale structures, the most relevant of which for inferring coalitions is community structure. Reflecting the organising principles of coalitions, community structure traditionally refers to groups of vertices that are more connected to one another than to vertices of other communities (Fortunato 2010). Demarcating such assortative communities from empirical networks is crucial for identifying coalition boundaries.

The ties from which various political phenomena emerge are however rarely constrained to singular contexts. Rather, the ties are multifaceted, shaped by past power struggles, and often occur across multiple relational contexts. Consider, for example, network-based studies of the U.S. Congress. Here, scholars have overwhelmingly focused on bill co-sponsorship networks (e.g., Fowler 2006; Zhang et al. 2008), in which ties are constructed from documented interactions in a highly formal context. Since congressional bills themselves are high-impact outcomes, legislative coalitions inferred from such ties are unsurprisingly predictive of American politics. But congressional members also interact in other ways, including sitting on the same committees (Porter, Mucha, Newman, and Warmbrand 2005), through "Dear Colleague" letters (Box-Steffensmeier, Christenson, and Craig 2019), holding joint press events (Desmarais, Moscardelli, Schaffner, and Kowal 2015), on social media (Chamberlain, Spezzano, Kettler, and Dit 2021), and through congressional staff exchange between offices (Montgomery and Nyhan 2017), all of which affect congressional behaviour.

Further, different contexts, through varying constraints and affordances, often yield different types of strategic interaction between actors, resulting in complex dependencies across contexts. Regarding the U.S. Congress, for example, Desmarais, Moscardelli, Schaffner, and Kowal (2015) find that joint press events between Congressional members are better at capturing intentional collaboration than patterns of bill co-sponsorship, as the latter tend to be constrained by partisanship. In this case, the two contexts essentially complement one another while serving different functions for actors in the system. Similar dependencies have been found across a range of political systems. Political campaigns in online and offline contexts complement one another in effective political communication (Freelon, Marwick, and Kreiss 2020), diplomats negotiate international conflict in both formal and informal contexts (Böhmelt 2010), conflict in the global system tends to result from belligerent actions in multiple contexts (Tir and Jasinski 2008), and actors engaging in policy-making are likely to harness multiple contexts to gather data and ideas (Leifeld and Schneider 2012).

It follows that while political actors are not constrained to act in singular contexts, ties in different contexts do not affect political phenomena independently of one another. To understand coalitions, both insights should be accounted for. Considering one context at one time would neglect the different functions that different contexts play for political actors. And, even if one would consider multiple contexts, neglecting that ties in one context hardly occur in isolation from other ties would neglect the potential for ties in one context also affecting ties in other contexts.

For studying how coalitions unfold across contexts, a traditional monolayer network approach falls short. Specifically, monolayer networks allow for comparing behaviour in different contexts but are unable to capture cross-context dependencies. Therefore, we resort to multilayer networks, an emerging paradigm in the study of complex systems, ranging from stellar constellations to the human connectome (Kivelä et al. 2014). As the name suggests, multilayer networks divide information into layers. Layers classify vertices into disjoint subsets, with each layer hosting its own network, which usefully serve to present different relational contexts, along both modal and temporal dimensions. How units and interactions are represented by the vertices and edges within and across layers comprise the multilayer network. Then, a multilayer conception of political alliances, *complex coalitions*, defines coalitions as groups that emerge from and evolve over time due to the relatively simple local ties among political actors that occur within and across relational contexts.

Methodologically, inferring complex coalitions is closely tied to the detection of community structure from empirical multilayer networks. To assist in describing the resulting structure, we identify four ideal-typical structures of complex coalitions (after Hanteer and Magnani 2020): pillar, semi-pillar, hierarchy, and overlap. Figure 1 depicts each as a section of a multilayer network with modal and temporal dimensions; vertices represent the same actors across layers, and each layer connects with its adjacent layers through vertices.

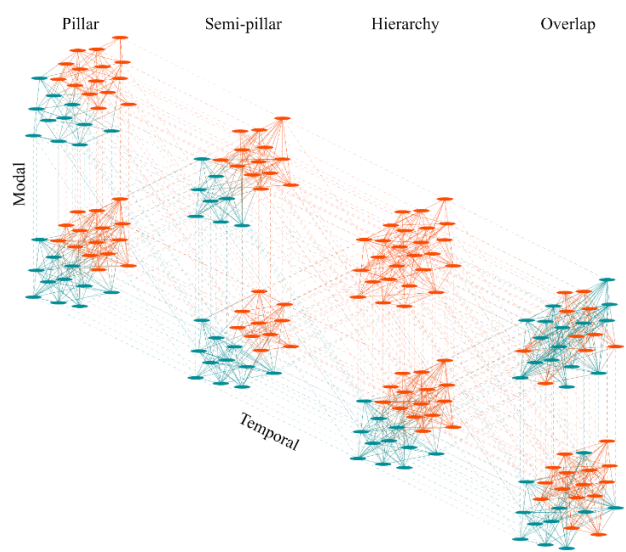

**Figure 1.** Ideal-typical structures of complex coalitions with modal and temporal axes depicting different relational contexts and evolution over time, respectively.

The *pillar structure* suggests that all members of coalitions participate in all relevant contexts along the modal axis and form roughly the same set of ties with the same actors in each, giving rise to the exact same coalition boundaries across contexts. For example, if the two contexts were bill co-sponsorship and joint press events in the US Congress, this would mean that all who co-sponsor a bill also tend to co-organise a press event. If some actors do not to participate in all contexts, but still form ties rather consistently in different context, the s*emi-pillar structure* captures such behaviour. To continue with the same example, this would mean that politicians still behave rather consistently across contexts, giving rise to the same coalition boundaries, but all who engage in co-sponsoring a bill do not co-organise a single press event. The *hierarchical* structure implies that coalitions that occur in one context splinter into smaller factions in another context. This would mean that one coalition of politicians that co-sponsored a set of bills on, say, gun control, which another coalition rejected, but the coalitions agreed to organise press events involving politicians from both sides. Alternatively, some context is more consensual, and segregation does not occur. In this case, politicians would have unanimously sponsored the bills, but held separate press events. In addition to the three ideal types, we consider the scenario of an *overlapping structure*, representing a realistic scenario of actors forming ties somewhat differently over time and/or across relational contexts. In Figure1, the temporal axis represents the evolution of the structure from pillar to overlap over time, while the cross-context edges reflect the modal and temporal dependencies.

More generally, the structures of complex coalitions depend on a) whether actors participate in all contexts and b) whether actors behave and form ties differently in different contexts. Observing any combination of the four structures would prompt questions about the meaning and function of the coalitions that emerge from different contexts. However, considering ties in different contexts is simply not enough to capture the notion of the different types of ties and coalitions rarely occurring is isolation from one another. For example, if one would be interested in some well-defined variety of coalitions that occur in some specific context, claiming coalition boundaries without accounting for the effect of cross-context dependencies risks misleading interpretations of coalition boundaries, and much of politics. To demonstrate how the insights from our theoretical-methodological complex coalitions framework play out in an empirical setting, we apply the framework to analyse complex coalitions over eight years of Finnish climate politics.

## 3 Case of Finnish climate politics

Under the complex coalitions framework, ties among political actors essentially occur across multiple relational contexts and exhibit complex dependencies. The different contexts also affect politics in different ways and provide different interpretations of various aspects of politics, such as power and polarisation. Determining the contexts that count most in different political realms calls for theoretical reasoning and empirical experimentation.

We illustrate our complex coalitions approach – from conceptualisation, to data preparation, to coalition inference, and finally to an analysis example – using climate politics in Finland as our case study. Our case is one where the characteristics of the contexts yield differentiation, and therefore tend to complement one another. The case involves both political contestation in different arenas and potential shifts in ties over time, which serve well to chart the implications of complex coalitions in contemporary democracies. Specifically, we look at how organisations engaging in climate politics interact within and across three contexts in the period of 2013-2020: resource coordination, legacy media discourse, and social media communication. The period is split by the 2016 Paris Agreement to demonstrate the role of temporal dynamics for complex coalition formation. Before discussing the three contexts, we describe the empirical setting.

### 3.1 Empirical setting

A key aspect of contemporary climate politics is about renegotiating the institutions that structure social and economic activities within societies (Aklin and Mildenberger 2020). Traditionally, in the consensual-corporatist political system of Finland, opposition to decisive climate action has been rooted in the coalition of aligned state, business, and labour interests. This status quo has mainly been challenged by administrative organisations, scientific authorities, and environmental organisations, although usually as two separate coalitions (Gronow and Ylä-Anttila 2019; Kukkonen and Ylä-Anttila 2020). In response to various pressures and opportunities, such as the Paris Agreement and the "green growth" paradigm, recent coalition governments have each raised the bar, assuming global climate leadership (Vikström, Mervaala, Kangas, and Lyytimäki 2023). Whereas climate politics currently manifest as competing articulations of the inherent contradictions of an agenda that pursues carbon neutrality, social justice, and economic growth all at once, coalition dynamics likely contributed to the recent upward shift in ambition.

### 3.2 Relevant contexts

Most network-based studies of coalitions focus on some expression of concrete coordination of action among political actors. However, the coalitions that emerge from ties in other contexts often play different functions, perhaps reinforcing, anticipating, or following such coordination. Specifically, the power that media exert on politics has grown considerably over the past years (Strömbäck 2008), which has increased the incentives for various actors to engage and interact with one another in different media contexts. Actors striving to affect politics are likely to act strategically to exploit the opportunities that different contexts provide, and form ties accordingly. In our case study, we consider coalitions emerging not only from joint resource coordination, but also from legacy media discourse and social media communication.

First, the need to consider resource coordination stems from the classic depiction of coalitions as groups of actors sharing political beliefs and coordinating actions (e.g., strategy work and resource mobilisation) over time to gain the upper hand in policy-making (Sabatier 1988). Such *advocacy coalitions* are deemed pervasive in governance as policymakers often rely on various resources from actors from various backgrounds. While belief homophily creates the necessary trust that enables the willing contribution of an actor to joint work (Kets and Sandroni 2019), the dominant advocacy coalition, emerging from such resource coordination, often gets to translate its core beliefs into concrete policy outcomes (Weible, Ingold, Nohrstedt, Henry, and Jenkins-Smith 2020). A consensual political system may encourage collaboration across coalition lines (Metz and Brandenberger n.d.), but coordination that lacks belief homophily is unlikely to give rise to advocacy coalitions.

Second, the importance of legacy media discourse arises from the notion of political discourse governing much of political activity through its structure; for example, ideas that are not part of the discourse are unlikely to materialise (Schmidt 2008). Hajer (1993) saw the discursive space in politics comprising *discourse coalitions*, the members of which interact through political statements that contribute to the creation of common discourses. Discourse coalitions, then, are groups of actors that share an interpretation of the political reality and try to impose the interpretation on others through media. A hegemonic discourse coalition often wields considerable agenda-setting power and widespread public support to leverage (Nelson 2004). Since content by professional journalists is still by far the most common source of information for the largest and the most affluent segment of population across democracies (Newman, Fletcher, Robertson, Eddy, et al. 2022), shaping discourse in legacy media ought to carry most utility for political actors.

Third, from the coalition perspective, social media communication has attracted much less attention than resource coordination or legacy media discourse. Social media platforms such as Twitter have become important arenas for political communication, letting actors to form ties by endorsing and spreading (i.e., retweeting) political content of other actors (Hong, Choi, and Kim 2019; Valenzuela, Correa, and Gil de Zúñiga 2018). As deliberate acts, endorsements serve to establish or maintain contacts, while providing an effective means to propagate information and impose ideas on others. We expect such behaviour to divide the online space into what we call *echo coalitions*. Unlike discourse coalitions, echo coalitions harness the expansive nature of social media to exert power over political information flow, power that most directly translates into agenda-setting power, potentially to the extent of driving discourse coalitions in legacy media (Gilardi, Gessler, Kubli, and Müller 2022).

Giving rise to advocacy coalitions, discourse coalitions, and echo coalitions, resource coordination, legacy media discourse, and social media communication, respectively, affect politics in different ways. The three contexts also further exhibit complex dependencies with one another, which lie at the heart of the complex coalitions approach.

### 3.3 Cross-context dependencies

To clearly illustrate the need to consider cross-context dependencies, we outline characteristics of the three contexts to emphasise the complementarity across the different types of coalitions emerging from different types of ties. We expect the contexts to differ along three dimensions (Table 1): transaction costs, public visibility, and external mediation.

**Table 1.** Plausible complementarity across resource coordination, legacy media discourse, and social media communication in national politics.

| | Resource coordination [advocacy coalitions] | Legacy media discourse [discourse coalitions] | Social media communication [echo coalitions] |
|---|---|---|---|
| Transaction costs | High | Moderate | Low |
| Public visibility | Low | High | Low |
| External mediation | Moderate | High | Moderate |

First, tie formation in each context entails different transaction costs (i.e., time, money, effort), while high-cost contexts force actors to consider the ties in which to invest (North 1990). Resource coordination, the emergence of which requires belief homophily and other resources to overcome inertia and materialise benefits, is one such context. The prospects of coordination are predicated upon the political opportunity structure, but high transaction costs imply that actors tend to prioritise ties that render most value (Fischer and Sciarini 2016). Once established, such collaboration often ingrains in practices and mutual expectations, and becomes increasingly costly to sever (Sedgwick and Jensen 2021; Siegel 2009). Political statements in legacy media also call for investments in expertise and campaigning to pass the publication threshold, but costs ought to be lower than for resource coordination, especially for prominent actors. Social media, however, has dramatically cut the transaction costs of forming ties, implying lower threshold for interacting with other actors, including with political opponents (Esteve Del Valle, Broersma, and Ponsioen 2022).

Second, public visibility of tie formation varies across contexts. When behaviour is highly visible to the public, actors have an incentive to consider public opinion. For example, while "underdogs" and actors who wish to expand the conflict are likely to pursue visibility, actors who either want to avoid drawing attention to an issue or whose goals deviate from the dominant public opinion may either adjust the message (cf., greenwashing) or tactically manage participation in high-visibility contexts, particularly if the political system does not encourage participation (Christiansen, Mach, and Varone 2018; Jensen and Seeberg 2020; Jones and McBeth 2010). Given the reach of legacy media, legacy media discourse exhibits high visibility. Although fast-paced communication on social media is likely attract less public attention, journalists increasingly report on such interactions, particularly during salient events (Figenschou and Fredheim 2020; McGregor 2019). Resource coordination ties, in turn, are not necessarily visible to the public at all.

Third, contexts differ with respect to the extent that participation depends on external mediation. If mediators are powerful, actors have less control over whether and when to participate, with whom to interact, and what are the outcomes of tie formation in the given context. Regarding collaboration, coalitions may have gatekeepers who regulate collaboration with other members (Gould and Fernandez 1989). Under such circumstances, collaboration depends not only on the beliefs and resources of the actor, but on the internal dynamics of the coalition (Olson 1965). In legacy media, journalists exert power over who gets to participate and how the message gets framed (Strömberg 2015). Actors must also consider the effects of journalistic norms on discourse that often serve to juxtapose rivals and emphasise conflict between the protagonists, especially when the debate heats up (Wahl-Jorgensen, Berry, Garcia-Blanco, Bennett, et al. 2017). Social media may quell participation through mechanisms such as witnessing incivility (Frimer et al. 2023), but platforms like Twitter do not predetermine terms of participation. The chances of retweeting political content are however mediated by algorithms (Huszár et al. 2022).

Taken together, we expect all three contexts to exhibit some degree of mutual reinforcement, but advocacy coalitions and echo coalitions emerging from resource coordination and social media communication, respectively, to exhibit higher degree of similarity in terms of community structure. We expect less participation in discourse coalitions (high public visibility and high external mediation) and less temporal shift in advocacy coalitions (high transaction costs). Consequently, complex coalitions in Finnish climate politics, before and after the Paris Agreement, are likely to combine semi-pillar and overlapping structures. Next, we outline the construction of the multilayer network from empirical data.

### 3.4 Multilayer network construction

Our multilayer representation of climate politics in Finland has six layers, which were constructed with data from surveys, newspapers, and Twitter, spanning four years before (2013-2016) and after (2017-2020) the Paris Agreement. For details on our data collection and processing, see page 1 in the appendix.

System boundaries were determined at the time of the first survey in 2014 and updated in 2020 before the second survey, constraining the layers before (*T0*) and after Paris (*T1*) to 96 and 103 organisations, respectively. Resource coordination (*Res*) was operationalised as the intersection of self-reported long-term political collaboration and response similarity over 10 most divisive political statements per survey. Response rates were sufficiently high (85%) for the inclusion of non-respondents. Thus, we chose to retain all organisations (no missing vertices) at the expense of not having data on collaboration among non-respondents (some missing edges). To operationalise reputational power (Fischer and Sciarini 2015), we aggregated indications of an organisation being influential in climate politics by all respondents.

Discursive alignment (*Dis*) was operationalised through agreement among organisations over 100 political statements in two prestige newspapers, following the Discourse Network Analysis procedure (Leifeld and Haunss 2012). To collect the (unquoted) retweets among organisations (*Com*), we applied the protocol by Chen, Kivelä, Malkamäki, Palosaari, et al. (2021) to identify the primary and potential secondary Twitter accounts of each organisation, and the accounts of its upper echelons. For example, from the Finnish Energy Authority, we have its primary account (@energiavirasto), secondary accounts such as its energy efficiency unit (@ekosuunnittelu), and the accounts of its main affiliates such as its deputy director (@velipekkasaajo). Retweets were filtered by relevant keywords and aggregated across the levels to capture the online behaviour of each organisation.

We had two temporal sets of three modal layers, all of which were undirected, unsigned, and weighted networks of high density [Res/T0: 26%, Res/T1: 30%, Dis/T0: 54%, Dis/T1: 46%, Com/T0: 36%, Com/T1: 48%]. However, as the data were generated through different processes, each with different measurement and different biases, there was high potential for non-salient edges. In addition, when most vertices connect to most other vertices, the underlying structure becomes difficult to parse. Thus, we extracted the latent binary structure of each layer according to estimates of error-variance in edge weights (Coscia and Neffke 2017). Edges with a p-value below 0.05 were kept, respectively yielding the following densities [16%, 17%, 10% 12%, 9%, 8%]. Figure 2 shows that all resulting layers are dissimilar in terms of edge overlap (Jaccard similarity) and degree distribution (Kendall correlation).

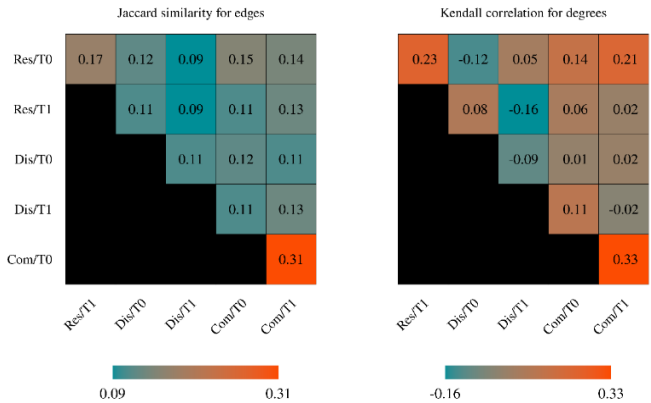

**Figure 2.** Similarity of layers for the intersection of vertices according to Jaccard similarity and Kendall correlation for edge overlap and degree distribution, respectively.

### 3.5 Coalition inference

To infer complex coalitions from multilayer networks, we combined two popular approaches to community detection: modularity maximisation and stochastic block modelling (SBM). Modularity measures the "quality" of a partition of a network into relatively dense communities (i.e., fewer edges between than within communities), with larger values corresponding to partitions with more edges within communities than what one would expect at random (Newman 2006). Of the various modularity-maximising algorithms, we used the Leiden algorithm that has been shown to perform well and return well-connected assortative communities (Traag, Waltman, and van Eck 2019), which also correspond to the organising principles of coalitions.

Maximising modularity for multilayer networks calls for specifying three sets of parameters, the appropriate values of which are imperative for uncovering meaningful community structure: *resolution* (γ, influences sizes of communities in each layer, Reichardt and Bornholdt 2006), *coupling* (ω, influences how much communities vary between layers, Mucha, Richardson, Macon, Porter, and Onnela 2010), and *layer weight* (β, influences how much community structure in one layer influences the structure in other layers, Pamfil, Howison, Lambiotte, and Porter 2019). As our three contexts represent clearly different dimensions of politics, we kept β uniform across layers, but estimating the optimal values for the remaining parameters (6 for γ, 15 for ω) was an open question, to which modularity could not provide an answer. Usually, the model is chosen qualitatively by exploring the community structures in the parameter space under the assumption that all sets of parameters are uniform across layers. To do so, one must relate modularity maximisation to generative models of networks with community structure (e.g., SBM), which allow for quantifying the likelihood of the network having been generated by the given partition.

While SBM has been generalised to multilayer settings, most of its variants restrict or prevent communities from varying across layers, are not limited to detecting assortative community structures, expect flat degree distributions, assume some given number of communities, or only apply to networks with ordinal layers (De Bacco, Power, Larremore, and Moore 2017; Jeub, Mahoney, Mucha, and Porter 2017; Peixoto and Rosvall 2017). Generalising work by Newman (2016) on monolayer networks, Pamfil, Howison, Lambiotte, and Porter (2019) showed that maximising modularity for multilayer networks is equivalent, given appropriate parametrisation, to maximising the log-likelihood of community assignments under the degree-corrected planted partition (stochastic block) model (PPM). Both are functions of multilayer partition (*g*) and comprise the same parameters, edges inside and outside communities (except for multiplicative and additive constants), with the same optimal community assignments.

Like multilayer modularity, the conditional log-likelihood $\mathbb{P}(g)$ under the multilayer PPM is defined simply as the sum over layer-specific and cross-layer terms. Given *g*, maximisation of either function without accounting for the latter set of terms corresponds to $\mathbb{P}_{intra}(g)$ – the monolayer approach. The multilayer approach entails devising cross-layer dependencies through a prior distribution on *g* that operationalises our belief of the cross-context dependencies before observing the data. Then, $\mathbb{P}_{inter}(g)$ depends only on ω and translates into the probability of vertices in one layer copying the community assignments from another layer (Pamfil, Howison, Lambiotte, and Porter 2019). However, calculating $\mathbb{P}_{inter}(g)$ for our empirical network with six layers remains computationally expensive as the process involves considering each permutation of layers for each vertex individually and summing over all terms to obtain $\mathbb{P}_{inter}(g)$ for each pair of layers.

To determine the optimal value for each parameter, we devised our own algorithm to consider the most complex scenario of adjacent layers and perform model selection by computing $\mathbb{P}(g)$ explicitly. The algorithm begins by computing log-likelihoods for various combinations of uniform γ and ω to select initial values, proceeds by modifying the value independently for each parameter, and updates the value if $\mathbb{P}(g)$ increases; the procedure repeats until an iteration is passed without updates. In addition to our algorithm being a novel implementation by enabling model selection and charting initial values (instead of arbitrary defaults or guesses), it also handles uneven participation in layers, implements consensus clustering to reconcile *g* over multiple runs of the stochastic Leiden algorithm (Peixoto 2021), and allows for incorporating prior information concerning the maximum number of communities in each layer (in our case, three).

Finally, the results for the best-fitting model were applied to analyse how the multilayer approach performs against the monolayer approach and changes our interpretations not only of the coalition structure, but of power and polarisation in politics. Concerning the latter, we employed two measures: reduced mutual information (RMI) and adaptive external-internal index (AEI). RMI quantifies the similarity of partitions as the amount of information that one obtains from one set of communities by observing another set of communities (Newman, Cantwell, and Young 2020). AEI quantifies network segregation as the ratio of edges within and between communities (Salloum, Chen, and Kivelä 2022). By appropriately handling varying numbers and sizes of communities, both measures represent recent extensions to popular measures in network science.

## 4 Results

Compared with the best-fitting monolayer model [$\mathbb{P}_{intra}(g)$ 2,595, $\mathbb{P}_{inter}(g)$ 0, $\mathbb{P}(g)$ 2,595], accounting for cross-context dependencies markedly improves model fit to data [$\mathbb{P}_{intra}(g)$ 2,566, $\mathbb{P}_{inter}(g)$ 115, $\mathbb{P}(g)$ 2,681]. The best-fitting multilayer model is $e^{86}$ = ~$10^{37}$ more likely than the monolayer model, based on which we may safely conclude that the former provides a substantially better fit.

Figure 3 illustrates the empirical network representing Finnish climate politics before and after the Paris Agreement, with coalition assignments according to the multilayer model. The monolayer approach finds three coalitions in the resource coordination context (Res) at T0 and two in T1, two coalitions in the legacy media discourse context (Dis) at both time points, and for the social media communication context (Com), two coalitions at T0 and three at T1. The multilayer discovers a third coalition in the media discourse context at T1. A change in the number of coalitions in one context, however, does not tell the whole story of the improvements of the multilayer approach on the coalition structure.

Figure 4 depicts the similarity of coalition structure for the intersection of actors in each pair of contexts. A notable change in RMI (monolayer to multilayer) occurs between the discourse contexts at T0 and T1 (-0.01 to 0.25). In the former, instead of misdiagnosing two coalitions of uneven size (monolayer; 57% and 11% of actors), we leverage cross-context correlations to enhance the "signal" of underlying coalition structure, reduce context-specific "noise", and evidence three coalitions (multilayer; 31%, 23%, and 14% of actors). Another change in RMI includes the increase between the discourse context at T0 and the resource coordination context at T1 (0.29 to 0.37), reinforcing the view that former anticipates the latter. In other words, actors begin interacting in legacy media discourse context at T0, which appears to have led to resource coordination at T1.

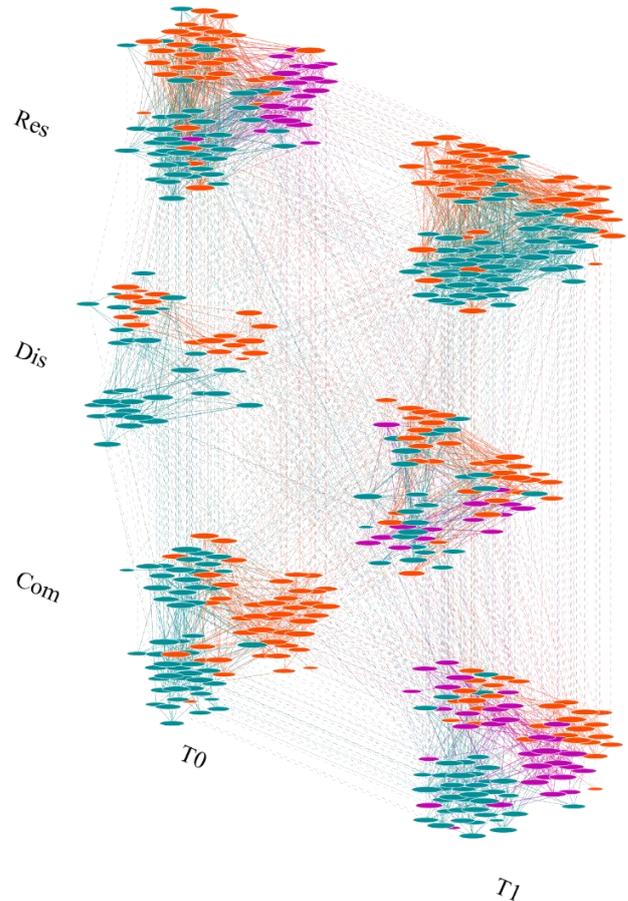

**Figure 3.** Complex coalitions in Finnish climate politics according to best-fitting multilayer model; vertex size reflects the number of ties of each actor in each context.

Small changes in RMI imply minor, but non-trivial corrections in the coalition assignments of organisations that seem to lie at the border of two coalitions in the monolayer solution. In the resource coordination context at T1, for example, the multilayer approach shows that the Finnish Innovation Fund, the seventh most powerful organisation in Finnish climate politics, falls into the competing coalition after accounting for cross-context dependencies. Generally, the structures are not notably more similar *within modal contexts over time* than *between modal contexts at one point in time*, despite accounting for cross-context dependencies.

The complex coalitions in Finnish climate politics, then, exhibit an overlapping structure, with uneven participation by actors across contexts. Both before and after Paris, the results support our expectation of the slightly closer resemblance of advocacy coalitions with echo coalitions ($RMI_{Res/T0-Com/T0}$ 0.14; $RMI_{Res/T1-Com/T1}$ 0.16) than with discourse coalitions ($RMI_{Res/T0-}$ $_{Dis/T0}$ 0.03; RMI$_{Res/T1-Dis/T1}$ 0.09). Apart from the similarity of coalition assignments, we find that legacy media discourse invites relatively few actors. Although the participation rate increases with the salience of climate affairs from 52% at T0 to 68% at T1, the respective rates for social media communication at 97% and 98% draw close to the rates for both resource coordination contexts at 100%. However, the share of reputational power that discourse coalitions capture (64% in at T0 and 84% at T1) exceeds the participation rates, implying that the key actors are present (Figure 5).

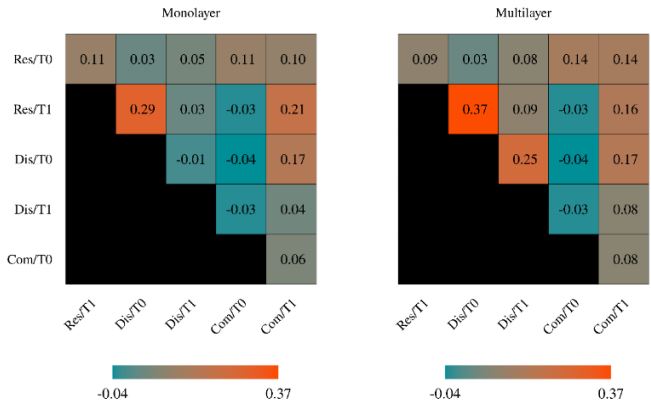

**Figure 4.** Similarity of partitions for the intersection of actors in each pair of contexts according to reduced mutual information; unlike in Figure 2, the colour scale applies to both sides.

A closer look at the structure reveals interesting dynamics. Before Paris, the three advocacy coalitions that engage in resource coordination represent smaller-scale environmental organisations and trade unions (purple), scientific organisations and some economic-industrial organisations (orange), and various economic-industrial and administrative organisations together with major political parties and larger-scale environmental organisations (turquoise). The orange and the turquoise manifest low polarisation across coalition lines (AEI 0.14) and capture 85% of reputational power, indicating rather consensual politics. The divide between echo coalitions in the social media, of which there are two due to the merger of the purple and the orange coalition, is much sharper (AEI 0.45). However, the discourse coalitions reshuffle the orange and the turquoise coalitions; scientific and environmental organisations unite under the orange discourse coalition (19% of actors, 18% of power), while most economic-industrial actors in the orange advocacy coalition shift to the turquoise discourse coalition (33% of actors, 46% of power). However, few economic-industrial actors participate in the discourse. After Paris, the two advocacy coalitions largely resemble the earlier discourse coalitions in terms of structure, but not in terms of polarisation (0.68 between discourse coalitions at T0 and 0.08 between advocacy coalitions at T1); the distribution of power also appears as uneven as before Paris. Both discourse coalitions and echo coalitions split into three; the purple rebounds when trade unions join hands with other major interest groups, while the purple echo coalition (32% of actors, 34% of power) differs from the purple discourse coalition (14% of actors, 17% of power) by inducing several political parties from the orange discourse coalition. The orange discourse coalition (that the monolayer approach failed to detect) is the only instance when a coalition with scientific and environmental organisations achieves the status of a hegemon, the discursive power of which likely laid its mark on the 2019-2023 government programme that promised plenty but achieved relatively little in terms of climate action.

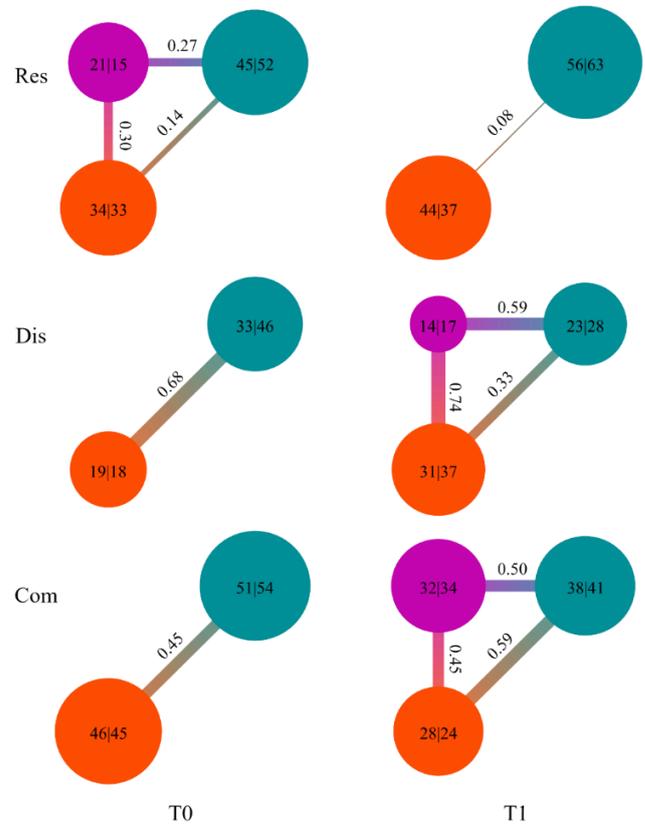

**Figure 5.** Power and polarisation in Finnish climate politics according to best-fitting multilayer model; vertex label indicates the relative shares of actors (left) and reputational power (right; vertex size) by coalition; higher values along edges reflect greater polarisation.

Overall, our interpretations of several key aspects of politics, such as power and polarisation, would have been misleading had we drawn our findings from a single context, without accounting for cross-context dependencies. Therefore, we conclude that political actors behave differently in different contexts, giving rise to different types of coalitions that also shape politics in different ways.

Had we observed ties only in legacy media, we would have underestimated the power of actors who never participate in the discourse. Had we mapped ties only on social media, we would have overestimated the clout of the orange coalition with environmental organisations at T0. Had we studied only resource coordination, we would have neglected an important dimension of most political processes: agenda-setting power. And, even if we would have recognised the need to consider ties in multiple contexts, had we not accounted for cross-context dependencies by using a multilayer approach, the results would have been less accurate and potentially misleading.

We theorised that differences in coalitions emerging from different contexts stem from the characteristics of each context along three dimensions: transaction costs, public visibility, and external mediation. Our key findings align with our expectations. For example, economic-industrial actors clearly participate less in legacy media, because of high public visibility and external mediation by journalists. The former carries unnecessary risks for actors for whom the corporatist system of Finland guarantees other means of influence, while the latter often emphasises conflict between the main protagonists from both sides (Christiansen, Mach, and Varone 2018; Wahl-Jorgensen, Berry, Garcia-Blanco, Bennett, and Cable 2017). Then, environmental organisations, the actor type commanding the fewest resources, are more prominent on social media because of low transaction costs and much lower level of external mediation. Communicating on social media is much less costly than coordinating resources with numerous other actors or mobilising campaigns that pass the publishing threshold in legacy media (Vesa, Poutanen, Sund, and Vehka 2022). The higher polarisation of both media contexts owes to public visibility and external mediation. In the least visible context of resource coordination, actors may subtly interact also with those whose goals are dissimilar, as encouraged by the consensus system of Finland, while journalists in legacy media follow norms according to which presenting opposing viewpoints to an issue is desirable (Wahl-Jorgensen, Berry, Garcia-Blanco, Bennett, and Cable 2017). Contrary to our expectation concerning the limited temporal shift in advocacy coalitions due to high transaction costs (Sedgwick and Jensen 2021), we observe an interesting cross-context dynamic suggesting that discourse coalitions (agenda-setting power) at T0 have translated into advocacy coalitions (policy-making power) at T1.

## 5 Conclusion

Coalitions are endemic to politics and contribute to various societally significant outcomes. Drawing from network science, we set out to advance understanding of coalitions by introducing *complex coalitions*, a theoretical-methodological framework that emphasises that coalitions emerge from the decisions of political actors to interact, but that the relevant ties are rarely limited to singular relational contexts. Rather, actors weigh the costs and benefits of interacting across multiple contexts, the coalitions in which also affect politics in different ways. Yet, ties in different contexts hardly occur independently of each other, which we argue must be appropriately accounted for when inferring coalitions. To do so, we contributed by implementing novel methodological tools to statistically infer complex coalitions from empirical multilayer networks.

To demonstrate the usefulness of our approach, we used data on ties among Finnish organisations engaging in climate politics across three specific contexts: resource coordination, legacy media discourse, and social media communication. We showed that our multilayer approach to coalition inference (i.e., interdependent contexts) improved model fit, and the empirical validity of findings, over the conventional monolayer approach (i.e., independent contexts). In addition to better fit, the multilayer approach refined the coalition assignments of certain key actors and uplifted meaningful coalitions that would have otherwise gone unnoticed. We may safely conclude that failure to consider cross-context dependencies when inferring coalitions of any type risks producing bias that may become large, especially when the structure in some context is fuzzy.

The results for our case study also address our question of whether actors behave differently in different contexts, leading to coalitions manifesting differently as well. Our results provide clear evidence for actors forming ties differently, and for the emergent coalitions manifesting differently in different modal and temporal contexts. The structure of complex coalitions in Finnish climate politics, 2013-2020, is overlapping, with uneven participation by actors in different contexts. For our three contexts, we observe closer resemblance for both coalition assignments and actor participation between advocacy coalitions (resource coordination) and echo coalitions (social media communication) than between advocacy coalitions and discourse coalitions (legacy media discourse). An answer to why contexts differ, we argue, lies in the different functions (agenda-setting and policy-making) and characteristics (transaction costs, public visibility, and external mediation) of each contexts. Ties that actors form across the three contexts are complementary, but hardly congruent channels for actors to enact coalitions.

It follows that by incorporating data on multiple types of ties we generate a more holistic understanding of coalitions, power, and polarisation in politics than by focussing on any singular context. The complex coalitions approach, encompassing a theoretical framework and a set of methodological tools, is directly generalisable to any political realm. However, the contexts that count in each realm may vary considerably.

For example, in international relations, the relevant types of ties almost certainly include trade and co-participation in intergovernmental organisations (Cranmer, Menninga, and Mucha 2015), but also co-signing of bilateral or multilateral treaties, co-voting in international institutions, and co-enacting of sanctions against common adversaries (Hafner-Burton, Kahler, and Montgomery 2009).

Areas of future work include extension of the complex coalitions framework with explicit hypotheses, including on issues such as spontaneous order (e.g., how stability emerges), nonlinearity (e.g., how does pace of change vary across contexts), feedback dynamics (e.g., how does change in one context anticipate change in another context), and factors contributing to different types of coalitions across political realms, systems, and cultures. Modern data labelling tools also carry potential to facilitate data collection and improve the quality of data on political ties (Benoit, Munger, and Spirling 2019), an issue comprising the most obvious practical limitation for applying the complex coalitions approach. However, following the rapid development of methods that account for complex dependencies and other complications, addressing questions concerning processes that give rise to emergent phenomena in politics, such as complex coalitions, is becoming increasingly feasible.

### Ethical review statement

Our work uses data from two surveys for Finnish political organisations. The original 2014 survey was repeated in 2020. Due to the research setting not meeting the requirements for conducting an ethical review by the Research Ethics Committee at the home institution of the authors who oversaw the survey design, we did not obtain an external review.

### Competing interests

The authors declare none.

### Code availability

All methodological tools and replication materials are hosted on *github.com/complexcoalition*.

### References

Aklin, Michaël, and Matto Mildenberger. 2020. "Prisoners of the Wrong Dilemma: Why Distributive Conflict, Not Collective Action, Characterizes the Politics of Climate Change." *Global Environmental Politics* 20(4): 4–27.

Aref, Samin, and Zachary P. Neal. 2021. "Identifying hidden coalitions in the US House of Representatives by optimally partitioning signed networks based on generalized balance." *Scientific Reports* 11(1): 19939.

Benoit, Kenneth, Kevin Munger, and Arthur Spirling. 2019. "Measuring and Explaining Political Sophistication through Textual Complexity." *American Journal of Political Science* 63(2): 491–508.

Böhmelt, Tobias. 2010. "The effectiveness of tracks of diplomacy strategies in third-party interventions." *Journal of Peace Research* 47(2): 167–178.

Boix, Carles, and Milan W. Svolik. 2013. "The Foundations of Limited Authoritarian Government: Institutions, Commitment, and Power-Sharing in Dictatorships." *Journal of Politics* 75(2): 300–316.

Box-Steffensmeier, Janet M., Dino P. Christenson, and Alison W. Craig. 2019. "Cue-Taking in Congress: Interest Group Signals from Dear Colleague Letters." *American Journal of Political Science* 63(1): 163–180.

Chamberlain, Joshua M., Francesca Spezzano, Jaclyn J. Kettler, and Bogdan Dit. 2021. "A Network Analysis of Twitter Interactions by Members of the U.S. Congress." *ACM Transactions on Social Computing* 4(1): 1–22.

Chen, Ted Hsuan Yun, Mikko Kivelä, Arttu Malkamäki, Esa Palosaari, et al. 2021. "Comparing Climate Change Policy Networks: Codebook for Twitter Account Collection V1.1." https://github.com/tedhchen/componMultilayer/blob/main/TwitterCodebook.pdf.

Chen, Ted Hsuan Yun. 2021. "Statistical inference for multilayer networks in political science." *Political Science Research and Methods* 9(2): 380–397.

Cheng, Huimin, Ye Wang, Ping Ma, and Amanda Murdie. 2021. "Communities and Brokers: How the Transnational Advocacy Network Simultaneously Provides Social Power and Exacerbates Global Inequalities." *International Studies Quarterly* 65(3): 724–738.

Christiansen, Peter Munk, André Mach, and Frédéric Varone. 2018. "How corporatist institutions shape the access of citizen groups to policy-makers: evidence from Denmark and Switzerland." *Journal of European Public Policy* 25(4): 526–545.

Coscia, Michele, and Frank M.H. Neffke. 2017. "Network Backboning with Noisy Data." *IEEE International Conference on Data Engineering* 33: 425–436.

Cranmer, Skyler J., Elizabeth J. Menninga, and Peter J. Mucha. 2015. "Kantian fractionalization predicts the conflict propensity of the international system." *Proceedings of the National Academy of Sciences* 112(38): 11812–11816.

De Bacco, Caterina, Eleanor A. Power, Daniel B. Larremore, and Cristopher Moore. 2017. "Community detection, link prediction, and layer interdependence in multilayer networks." *Physical Review E* 95(4): 042317.

Desmarais, Bruce A., Vincent G. Moscardelli, Brian F. Schaffner, and Michael S. Kowal. 2015. "Measuring legislative collaboration: The Senate press events network." *Social Networks* 40: 43–54.

Desmarais, Bruce A., Raymond J. La Raja, and Michael S. Kowal. 2015. "The Fates of Challengers in U.S. House Elections: The Role of Extended Party Networks in Supporting Candidates and Shaping Electoral Outcomes." *American Journal of Political Science* 59(1): 194–211.

Ecker, Alejandro, and Thomas M. Meyer. 2020. "Coalition Bargaining Duration in Multiparty Democracies." *British Journal of Political Science* 50(1): 261–280.

Esteve Del Valle, Marc, Marcel Broersma, and Arnout Ponsioen. 2022. "Political Interaction Beyond Party Lines: Communication Ties and Party Polarization in Parliamentary Twitter Networks." *Social Science Computer Review* 40(3): 736–755.

Farrell, Justin. 2016. "Network structure and influence of the climate change counter-movement." *Nature Climate Change* 6(4): 370–374.

Figenschou, Tine Ustad, and Nanna Alida Fredheim. 2020. "Interest groups on social media: Four forms of networked advocacy." *Journal of Public Affairs* 20(2): e2012.

Fischer, Manuel, and Pascal Sciarini. 2016. "Drivers of Collaboration in Political Decision Making: A Cross-Sector Perspective." *Journal of Politics* 78(1): 63–74.

Fischer, Manuel, and Pascal Sciarini. 2015. "Unpacking reputational power: Intended and unintended determinants of the assessment of actors' power." *Social Networks* 42: 60–71.

Fortunato, David, and Randolph T. Stevenson. 2013. "Perceptions of Partisan Ideologies: The Effect of Coalition

Participation." *American Journal of Political Science* 57(2): 459–477.

Fortunato, Santo. 2010. "Community detection in graphs." *Physics Reports* 486(3): 75–174.

Fowler, James H. 2006. "Connecting the Congress: A Study of Cosponsorship Networks." *Political Analysis* 14(4): 456–487.

Freelon, Deen, Alice Marwick, and Daniel Kreiss. 2020. "False equivalencies: Online activism from left to right." *Science* 369(6508): 1197–1201.

Frimer, Jeremy A. et al. 2023. "Incivility Is Rising Among American Politicians on Twitter." *Social Psychological and Personality Science* 14(2): 259–269.

Gilardi, Fabrizio, Theresa Gessler, Maël Kubli, and Stefan Müller. 2022. "Social Media and Political Agenda Setting." *Political Communication* 39(1): 39–60.

Gould, Roger V., and Roberto M. Fernandez. 1989. "Structures of Mediation: A Formal Approach to Brokerage in Transaction Networks." *Sociological Methodology* 19: 89–126.

Gronow, Antti, and Tuomas Ylä-Anttila. 2019. "Cooptation of ENGOs or Treadmill of Production? Advocacy Coalitions and Climate Change Policy in Finland." *Policy Studies Journal* 47(4): 860–881.

Hafner-Burton, Emilie M., Miles Kahler, and Alexander H. Montgomery. 2009. "Network Analysis for International Relations." *International Organization* 63(3): 559–592.

Hagen, Achim, and Jan Schneider. 2021. "Trade sanctions and the stability of climate coalitions." *Journal of Environmental Economics and Management* 109: 102504.

Hajer, Maarten. 1993. "Discourse Coalitions and the Institutionalization of Practice." In *The Argumentative Turn in Policy Analysis and Planning*, Durham, NC: Duke University Press, p. 43–76.

Hanteer, Obaida, and Matteo Magnani. 2020. "Unspoken Assumptions in Multi-layer Modularity maximization." *Scientific Reports* 10(1): 11053.

Hong, Sounman, Haneul Choi, and Taek Kyu Kim. 2019. "Why Do Politicians Tweet? Extremists, Underdogs, and Opposing Parties as Political Tweeters." *Policy & Internet* 11(3): 305–323.

Huszár, Ferenc et al. 2022. "Algorithmic amplification of politics on Twitter." *Proceedings of the National Academy of Sciences* 119(1): e2025334119.

Jensen, Carsten, and Henrik Bech Seeberg. 2020. "On the enemy's turf: exploring the link between macro- and micro-framing in interest group communication." *Journal of European Public Policy* 27(7): 1054–1073.

Jeub, Lucas G. S., Michael W. Mahoney, Peter J. Mucha, and Mason A. Porter. 2017. "A local perspective on community structure in multilayer networks." *Network Science* 5(2): 144–163.

Johan, Hellström, and Jonas Lindahl. 2021. "Sweden: The Rise and Fall of Bloc Politics." In *Coalition Governance in Western Europe*, eds. Torbjörn Bergman, Hanna Back, and Johan Hellström. Oxford: Oxford University Press, p. 574–610.

Jones, Michael D., and Mark K. McBeth. 2010. "A Narrative Policy Framework: Clear Enough to Be Wrong?" *Policy Studies Journal* 38(2): 329–353.

Karol, David. 2009. *Party Position Change in American Politics: Coalition Management*. Cambridge: Cambridge University Press.

Kets, Willemien, and Alvaro Sandroni. 2019. "A belief-based theory of homophily." *Games and Economic Behavior* 115: 410–435.

Kivelä, Mikko et al. 2014. "Multilayer networks." *Journal of Complex Networks* 2(3): 203–271.

Kukkonen, Anna, and Tuomas Ylä-Anttila. 2020. "The Science-Policy Interface as a Discourse Network: Finland's Climate Change Policy 2002–2015." *Politics and Governance* 8(2): 200–214.

Leeds, Brett Ashley. 2003. "Do Alliances Deter Aggression? The Influence of Military Alliances on the Initiation of Militarized Interstate Disputes." *American Journal of Political Science* 47(3): 427–439.

Leifeld, Philip, and Sebastian Haunss. 2012. "Political discourse networks and the conflict over software patents in Europe." *European Journal of Political Research* 51(3): 382–409.

Leifeld, Philip, and Volker Schneider. 2012. "Information Exchange in Policy Networks." *American Journal of Political Science* 56(3): 731–744.

Mahoney, Christine, and Frank R. Baumgartner. 2015. "Partners in Advocacy: Lobbyists and Government Officials in Washington." *Journal of Politics* 77(1): 202–215.

Malkamäki, Arttu, Tuomas Ylä-Anttila, Maria Brockhaus, Anne Toppinen, et al. 2021. "Unity in diversity? When advocacy coalitions and policy beliefs grow trees in South Africa." *Land Use Policy* 102: 105283.

McClendon, Gwyneth H. 2014. "Social Esteem and Participation in Contentious Politics: A Field Experiment at an LGBT Pride Rally." *American Journal of Political Science* 58(2): 279–290.

McGregor, Shannon C. 2019. "Social media as public opinion: How journalists use social media to represent public opinion." *Journalism* 20(8): 1070–1086.

Metz, Florence, and Laurence Brandenberger. "Policy Networks Across Political Systems." *American Journal of Political Science*.

Montgomery, Jacob M., and Brendan Nyhan. 2017. "The Effects of Congressional Staff Networks in the US House of Representatives." *Journal of Politics* 79(3): 745–761.

Morrison, Margaret. 2015. "Why Is More Different?" In *Why More Is Different: Philosophical Issues in Condensed Matter Physics and Complex Systems*, eds. Brigitte Falkenburg and Margaret Morrison. Berlin: Springer, p. 91–114.

Mucha, Peter J., Thomas Richardson, Kevin Macon, Mason A. Porter, et al. 2010. "Community Structure in Time-Dependent, Multiscale, and Multiplex Networks." *Science* 328(5980): 876–878.

Nelson, Thomas E. 2004. "Policy Goals, Public Rhetoric, and Political Attitudes." *Journal of Politics* 66(2): 581–605.

Newman, M. E. J. 2016. "Equivalence between modularity optimization and maximum likelihood methods for community detection." *Physical Review E* 94(5): 052315.

Newman, M. E. J. 2006. "Modularity and community structure in networks." *Proceedings of the National Academy of Sciences* 103(23): 8577–8582.

Newman, M. E. J., George T. Cantwell, and Jean-Gabriel Young. 2020. "Improved mutual information measure for clustering, classification, and community detection." *Physical Review E* 101(4): 042304.

Newman, Nic, Richard Fletcher, Craig T. Robertson, Kirsten Eddy, et al. 2022. *Reuters Institute Digital News Report*. Oxford: Reuters Institute for the Study of Journalism, University of Oxford. https://reutersinstitute.politics.ox.ac.uk/digital-news-report/2021.

North, Douglass C. 1990. "A Transaction Cost Theory of Politics." *Journal of Theoretical Politics* 2(4): 355–367.

Olson, Mancur. 1965. *The Logic of Collective Action: Public Goods and the Theory of Groups, Second Printing with a New Preface and Appendix*. Cambridge, MA: Harvard University Press.

Pamfil, A. Roxana, Sam D. Howison, Renaud Lambiotte, and Mason A. Porter. 2019. "Relating Modularity Maximization and Stochastic Block Models in Multilayer Networks." *SIAM Journal on Mathematics of Data Science* 1(4): 667–698.

Park, Hyung. 2008. "Forming Coalitions: A Network-Theoretic Approach to the Contemporary South Korean Environmental Movement." *Mobilization: An International Quarterly* 13(1): 99–114.

Peixoto, Tiago P. 2021. "Revealing Consensus and Dissensus between Network Partitions." *Physical Review X* 11(2): 021003.

Peixoto, Tiago P., and Martin Rosvall. 2017. "Modelling sequences and temporal networks with dynamic community structures." *Nature Communications* 8(1): 582.

Porter, Mason A., Peter J. Mucha, M. E. J. Newman, and Casey M. Warmbrand. 2005. "A network analysis of committees in the U.S. House of Representatives." *Proceedings of the National Academy of Sciences* 102(20): 7057–7062.

Reichardt, Jörg, and Stefan Bornholdt. 2006. "Statistical mechanics of community detection." *Physical Review E* 74(1): 016110.

Roberts, Cynthia. 2019. "The BRICS in the Era of Renewed Great Power Competition." *Strategic Analysis* 43(6): 469–486.

Sabatier, Paul A. 1988. "An advocacy coalition framework of policy change and the role of policy-oriented learning therein." *Policy Sciences* 21(2): 129–168.

Sabatier, Paul A. 1998. "The advocacy coalition framework: revisions and relevance for Europe." *Journal of European Public Policy* 5(1): 98–130.

Salloum, Ali, Ted Hsuan Yun Chen, and Mikko Kivelä. 2022. "Separating Polarization from Noise: Comparison and Normalization of Structural Polarization Measures." *Proceedings of the ACM on Human-Computer Interaction* 6: 1–33.

Schmidt, Vivien A. 2008. "Discursive Institutionalism: The Explanatory Power of Ideas and Discourse." *Annual Review of Political Science* 11(1): 303–326.

Sedgwick, Donna, and Laura S Jensen. 2021. "Path Dependence and the Roots of Interorganizational Relationship Challenges." *Perspectives on Public Management and Governance* 4(1): 47–62.

Siegel, David A. 2009. "Social Networks and Collective Action." *American Journal of Political Science* 53(1): 122–138.

Strömbäck, Jesper. 2008. "Four Phases of Mediatization: An Analysis of the Mediatization of Politics." *International Journal of Press/Politics* 13(3): 228–246.

Strömberg, David. 2015. "Media and Politics." *Annual Review of Economics* 7(1): 173–205.

Tir, Jaroslav, and Michael Jasinski. 2008. "Domestic-Level Diversionary Theory of War: Targeting Ethnic Minorities." *Journal of Conflict Resolution* 52(5): 641–664.

Traag, V. A., L. Waltman, and N. J. van Eck. 2019. "From Louvain to Leiden: guaranteeing well-connected communities." *Scientific Reports* 9(1): 5233.

Valenzuela, Sebastián, Teresa Correa, and Homero Gil de Zúñiga. 2018. "Ties, Likes, and Tweets: Using Strong and Weak Ties to Explain Differences in Protest Participation Across Facebook and Twitter Use." *Political Communication* 35(1): 117–134.

Van Dyke, Nella, and Bryan Amos. 2017. "Social movement coalitions: Formation, longevity, and success." *Sociology Compass* 11(7): e12489.

Vesa, Juho, Petro Poutanen, Reijo Sund, and Mika Vehka. 2022. "An effective 'weapon' for the weak? Digital media and interest groups' media success." *Information, Communication & Society* 25(2): 258–277.

Vikström, Suvi, Erkki Mervaala, Hanna-Liisa Kangas, and Jari Lyytimäki. 2023. "Framing climate futures: the media representations of climate and energy policies in Finnish broadcasting company news." *Journal of Integrative Environmental Sciences* 20(1): 2178464.

Wahl-Jorgensen, Karin, Mike Berry, Iñaki Garcia-Blanco, Lucy Bennett, and Jonathan Cable. 2017. "Rethinking balance and impartiality in journalism? How the BBC attempted and failed to change the paradigm." *Journalism* 18(7): 781–800.

Weible, Christopher M., Karin Ingold, Daniel Nohrstedt, Adam Douglas Henry, and Hank C. Jenkins-Smith. 2020. "Sharpening Advocacy Coalitions." *Policy Studies Journal* 48(4): 1054–1081.

Weitsman, Patricia A. 1997. "Intimate enemies: The politics of peacetime alliances." *Security Studies* 7(1): 156–193.

Zhang, Yan et al. 2008. "Community structure in Congressional cosponsorship networks." *Physica A: Statistical Mechanics and its Applications* 387(7): 1705–1712.